\documentclass[a4paper,fleqn]{cas-dc}

\usepackage[sort&compress,numbers]{natbib}
\usepackage[utf8]{inputenc}
\usepackage{textcomp}
\DeclareUnicodeCharacter{22EF}{\dots}
\DeclareUnicodeCharacter{0302}{\^}

\usepackage{soul}

\def\tsc#1{\csdef{#1}{\textsc{\lowercase{#1}}\xspace}}
\tsc{WGM}
\tsc{QE}
\tsc{EP}
\tsc{PMS}
\tsc{BEC}
\tsc{DE}

\begin{document}
\let\WriteBookmarks\relax
\def\floatpagepagefraction{1}
\def\textpagefraction{.001}
\shorttitle{From 2D to 1D in $\beta$-Naphthyne}
\shortauthors{Laranjeira et~al.}

\title [mode = title]{From 2D to 1D in $\beta$-Naphthyne: A Porous Carbon Allotrope Merging Graphyne and Naphthylene}

\author[1]{José A. S. Laranjeira}
\affiliation[1]{
organization={Modeling and Molecular Simulation Group},
addressline={São Paulo State University (UNESP), School of Sciences}, 
city={Bauru},
postcode={17033-360}, 
state={SP},
country={Brazil}}
\cormark[1]
\cortext[cor1]{Corresponding author}
\credit{Conceptualization of this study, Methodology, Review and editing, Investigation, Formal analysis, Writing -- review \& editing, Writing -- original draft}
\author[2,3]{K. A. L. Lima}
\affiliation[2]{
organization={Institute of Physics},
addressline={University of Brasília}, 
city={Brasília },
postcode={70910‑900}, 
state={DF},
country={Brazil}}
\affiliation[3]{
organization={Computational Materials Laboratory, LCCMat, Institute of Physics},
addressline={University of Brasília}, 
city={Brasília },
postcode={70910‑900}, 
state={DF},
country={Brazil}}
\credit{Conceptualization of this study, Methodology, Review and editing, Investigation, Formal analysis, Writing -- review \& editing, Writing -- original draft}
\author[1]{Nicolas F. Martins}
\credit{Conceptualization of this study, Methodology, Review and editing, Investigation, Formal analysis, Writing -- review \& editing, Writing -- original draft}
\author[4]{Marcelo L. P. Junior}
\affiliation[4]{
organization={University of Bras{\'{i}}lia, College of Technology, Department of Electrical Engineering}, 
city={Brasília},
postcode={70910-900}, 
state={DF},
country={Brazil}}
\credit{Conceptualization of this study, Methodology, Review and editing, Investigation, Formal analysis, Writing -- review \& editing, Writing -- original draft}
\author[2,3]{L.A. Ribeiro Junior}
\credit{Conceptualization of this study, Methodology, Review and editing, Investigation, Formal analysis, Writing -- review \& editing, Writing -- original draft}
\author[1]{Julio R. Sambrano}
\credit{Conceptualization of this study, Methodology, Review and editing, Investigation, Formal analysis, Writing -- review \& editing, Writing -- original draft}

\begin{abstract}
Two-dimensional (2D) carbon-based materials have attracted considerable interest due to their diverse structural and electronic properties, making them ideal for next-generation flat electronics. Among these materials, metallic-like porous structures offer advantages such as tunable charge transport and high surface area, which are essential for energy storage applications. In this study, we introduce $\beta$-naphthyne, a novel 2D carbon allotrope composed of naphthyl units interconnected by octagonal rings. First-principles calculations confirm its dynamic and thermal stability, demonstrating its theoretical feasibility. Our results confirm its energetic, dynamic, and mechanical stability and its metallic nature. Furthermore, we demonstrate that Young’s modulus ranges from 46.60 N/m to 164.47 N/m, indicating an anisotropic mechanical response. Optical analysis reveals absorption activity in the infrared (IR) and ultraviolet (UV) regions. The 1D structures were also analyzed, revealing a Dirac cone and a transition from metallic to semiconducting behavior. These findings establish $\beta$-naphthyne as a promising material for energy storage and optoelectronic technologies.

\end{abstract}

% \includegraphics{figs/grabs.pdf}

%\item $\beta$-naphthyne is a novel porous 2D carbon allotrope.
%\item It features metallic-like behavior, ideal for energy storage with tunable charge transport.
%\item The structure consists of naphthyl units connected by octagonal rings.
%\item High mechanical anisotropy is observed in elasticity calculations.
%\item Strong optical absorption in IR and UV suggests optoelectronic potential.


\maketitle

\section{Introduction}

Two-dimensional (2D) carbon-based materials have garnered significant attention due to their mechanical, electronic, and optical properties, making them promising candidates for energy storage and nanoelectronics applications \cite{tiwari2016magical,nasir2018carbon,jana2021emerging}. Among them, porous 2D carbon allotropes offer additional advantages while maintaining a high surface-to-volume ratio, exhibiting tunable electronic structures and enhanced ion diffusion, which are crucial for hydrogen storage, supercapacitors, and metal-ion batteries \cite{zhang2014recent,zheng2015two,wu2022exploring,wu20202020,zhang2019recent,huang2020chemistry}. The incorporation of ordered nanopores into carbon frameworks can introduce novel electronic behaviors, such as Dirac cone semimetallicity, paving the way for the development of high-performance energy materials \cite{xu2012porous,huang2020chemistry,guirguis2020applications}. In this context, metallic-like porous 2D carbon allotropes can facilitate rapid charge transport while maintaining mechanical stability, essential properties for advanced energy storage applications.

Several porous carbon allotropes have been synthesized or theoretically predicted in recent years, each exhibiting distinct structural and electronic characteristics \cite{zhou2014general,zhang2014recent,aliev2025planar,desyatkin2022scalable,fang20222d,cavalheiro2024can,kharisov2023density}. For instance, graphdiyne, a synthesized material featuring acetylenic linkages, exhibits enhanced electrochemical performance due to its high charge carrier mobility and tunable bandgap \cite{gao2019graphdiyne}. Other experimentally synthesized materials include the biphenylene network, which was recently fabricated via on-surface synthesis and displays metallic behavior, holding potential for nanoelectronic applications \cite{fan2021biphenylene}. In contrast, several computationally predicted structures \cite{liu2012structural,zhang2015penta,tromer2023mechanical,junior2023irida,majidi2024irida,lima2025th,lima2024dodecanophene,song2013graphenylene,laranjeira2024graphenyldiene} have been characterized, revealing distinct structural, electronic, optical, and mechanical properties. However, the experimental realization of some of these structures remains a challenge \cite{ewels2015predicting}. Nevertheless, theoretical studies are essential for guiding experimental synthesis by predicting stability, mechanical response, and electronic characteristics, enabling researchers to design novel materials with desirable functionalities before fabrication \cite{jain2016computational,ramprasad2017machine}.

Building upon recent advances in 2D carbon-based materials for surface electronics, we introduce Naphthylene-$\beta$ Graphyne ($\beta$-naphthyne) as a novel porous 2D carbon allotrope a novel porous 2D carbon allotrope characterized by naphthalene molecules covalently linked through acetylenic groups at all their dehydrogenated terminations, forming a rectangular symmetry. $\beta$-naphthyne is a graphyne-like version of $\beta$-naphthylene, previously proposed by Paz \textit{et al.}, in which the allotrope is predicted from the naphthalene precursor in its so-called $\beta$ phase, configuring a 2D carbon allotrope with 4-, 6-, and 10-membered rings \cite{ALVARESPAZ2019792}. In the structure proposed here, the linkage between naphthalene molecules is replaced by acetylenic groups, transforming the originally sp$^2$-bonded system into a hybrid sp-sp$^2$ configuration, now exhibiting 6- and 8-membered rings, along with nanopores consisting of 18 carbon-membered units. 

This structural design enhances electronic delocalization, offering a promising framework for energy storage applications. However, regarding the challenges of synthesizing such materials, naphthalene units no longer serve as a precursor for a potential synthesis of $\beta$-naphthyne. Instead, Octamethylnaphthalene (C$_{18}$H$_{24}$), synthesized in 1972 by Hart and Oku \cite{hart1972synthesis}, emerges as a viable structural building block for bottom-up synthesis strategies, such as surface-assisted polymerization or templated chemical vapor deposition (CVD), providing a feasible route for future syntheses of $\beta$-naphthyne \cite{han2021surface,ariga2023materials,clair2019controlling}.

%Unlike other 2D carbon allotropes that exhibit Dirac points without the hexagonal symmetry of graphene \cite{xu2014two, wang2015rare, liu2012structural}, $\beta$-naphthyne is a novel Dirac cone material with high charge carrier mobility. Given the increasing interest in porous carbon materials for energy applications, this study provides key insights into the properties of $\beta$-naphthyne. This work investigates its structural stability, electronic properties, and mechanical response through first-principles calculations. 

This study presents the novel $\beta$-naphthyne structure, investigated through first-principles calculations. Using density functional theory (DFT), we explored its electronic properties, including band structures, electron localization functions (ELF), and Bader charge distribution. The thermal and dynamic stabilities were verified using \textit{ ab initio} molecular dynamics (AIMD) simulations. A comprehensive mechanical evaluation was conducted by analyzing elastic constants, with stability confirmed via the Born-Huang criteria. In addition, the effect of 1D confinement was evaluated by analysis of $\beta$-naphthyne nanoribbons. The findings reported here not only highlight the $\beta$-naphthyne as a 2D reliable carbon allotrope but also suggest its potential as a platform for diverse applications, such as energy storage, gas detection, and electronics.

\section{Methodology}

To explore the structural, electronic, mechanical, and optical properties of $\beta$-naphthyne, we conducted first-principles simulations based on DFT \cite{hohenberg1964inhomogeneous}. The computational framework was implemented using the Vienna Ab Initio Simulation Package (VASP) \cite{kresse1999ultrasoft}. For electron exchange and correlation effects, we employed the generalized gradient approximation (GGA) following the Perdew-Burke-Ernzerhof (PBE) functional \cite{perdew1996generalized}. The interaction between valence and core electrons was treated using the projector-augmented wave (PAW) method \cite{kresse1996efficient}, which enhances computational efficiency while preserving high precision.

The kinetic energy cutoff for the plane-wave basis set was set to 520 eV, ensuring an accurate representation of electronic states. The structural optimizations were performed using the conjugate gradient method until the force on each atom was lower than 0.01 eV/\r{A}, and the total energy change between successive steps was below 10$^{-5}$ eV. A vacuum spacing of 15 \r{A} along the out-of-plane direction was introduced to prevent interactions between periodic images. The Brillouin zone was sampled using a $\Gamma$-centered Monkhorst-Pack \textbf{k}-point mesh of $3\times 3\times 1$ ($18\times 18\times 1$) for monolayers and $3\times 1 \times 1$ ($18\times 1\times 1$) for nanoribbons for optimization (density of states) calculations. Dispersion interactions were included using Grimme's DFT-D2 method \cite{grimme2006semiempirical}, accounting for van der Waals forces to accurately describe interatomic interactions in porous carbon frameworks.

To evaluate the dynamical stability of $\beta$-naphthyne, phonon dispersion calculations were performed using density functional perturbation theory (DFPT) as implemented in the Phonopy package \cite{togo2015first}. Additionally, \textit{ab initio} molecular dynamics (AIMD) simulations were conducted at 300 K for 5 ps in the NVT ensemble, employing the Nos\'e-Hoover thermostat \cite{nose1984unified,hoover1985canonical} to assess its thermal robustness.

Electronic properties were examined through band structure and density of states calculations. The band structure was computed along high-symmetry paths of the Brillouin zone to reveal Dirac cone features. At the same time, the projected density of states was used to analyze orbital contributions to electronic states. The first-order resonant Raman spectra at various laser energies were also estimated. 

The optical properties were investigated through calculations of the frequency-dependent dielectric function, from which absorption coefficients and reflectance spectra were derived \cite{moore2020optical}, providing insight into the potential optoelectronic applications of $\beta$-naphthyne.

The thermodynamic stability of $\beta$-naphthyne was analyzed by its cohesive energy (\(E_{\text{coh}}\)):

\begin{equation}
E_{\text{coh}} = \frac{E_{\text{$\beta$-naphthyne}} - \sum_i n_iE_i}{\sum_i n_i},
\end{equation}

where \(E_{\text{$\beta$-naphthyne}}\) is the total energy of the structure, \(E_i\) represents the energy of isolated C atoms, and \(n_i\) is the atom count of each element. The same approach was used to obtain the $E_{\text{coh}}$ for other 2D structures discussed in this work.

To analyze the vibrational properties of $\beta$-naphthyne, the coupled perturbed HF/Kohn–Sham algorithm \cite{ferrero2008coupled} implemented in the computational package CRYSTAL17 \cite{dovesi2005crystal} was used. The calculations in CRYSTAL17 were performed using triple-zeta valence with polarization (TZVP) basis set \cite{vilela2019bsse} together with the B3LYP functional. The structure was optimized by checking the root mean square (RMS) and the absolute value of the largest component of both the gradients and the estimated displacements. The convergence criteria used in the optimization for RMS and the largest component for gradient (0.00030 a.u. and 0.00045 a.u.) and displacement (0.00120 a.u. and 0.00180 a.u.), respectively. The reciprocal space was sampled using Pack-Monkhost and Gilat nets with sublattice and a shrinking factor of 12.

\section{Results}
\subsection{Structure and Stability}

\begin{figure*}[pos=!htb]
    \centering
    \includegraphics[width=1\linewidth]{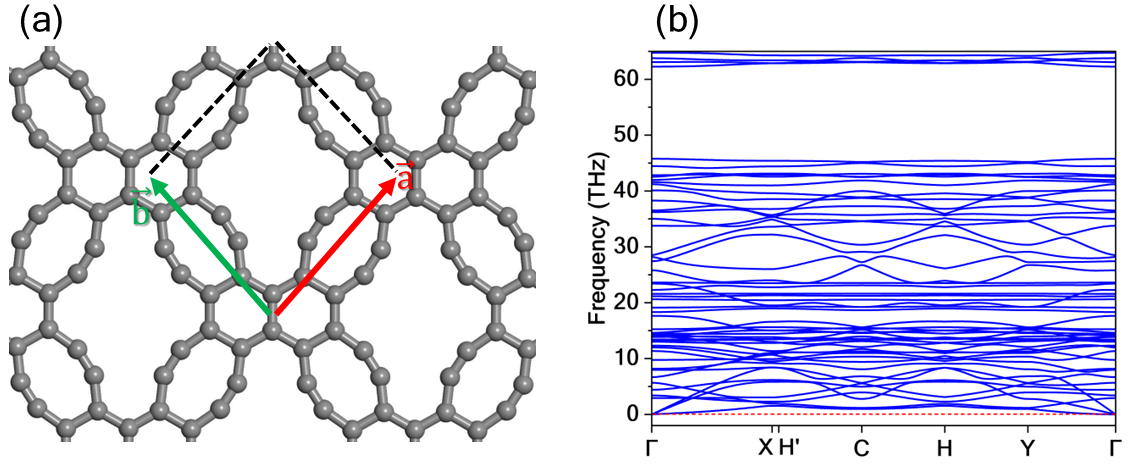}
    \caption{(a) Top view of the optimized atomic structure of $\beta$-naphthyne. The unit cell has primitive lattice vectors $a=b=7.82$ \r{A}. (b) Phonon dispersion spectrum of $\beta$-naphthyne along high-symmetry paths of the Brillouin zone.}
    \label{fig:system}
\end{figure*}

$\beta$-naphthyne is characterized by a rectangular conventional unit cell that belongs to the $Cmm$ (no. 65) space group, with lattice parameters $a=10.74$ \r{A} and $b=11.72$ \r{A} (see Fig. \ref{fig:system}). The primitive cell comprises lattice parameters $a=b=7.82$ \r{A}, $\alpha=\beta=90^\circ$, and $\gamma = 93.17^\circ$. The conventional unit cell consists of five crystallographically distinct carbon atoms located in C1 (0.764, 0.378, 0.500), C2 (0.359, 0.181, 0.500), C3 (0.566, 0.282, 0.500), C4 (0.367, 0.392, 0.500), and C5 (0.932, 0.000, 0.500). This structure forms a 2D carbon framework comprising 6-8 carbon-membered rings, where octagonal rings interconnect naphthyl fragments. For comparison purposes, Table \ref{tab:properties_comparison} details the lattice parameters ($a$, $b$, $\alpha = \beta$, and $\gamma$), space group (SPG), band gap energy ($E_\text{gap}$), and cohesive energy ($E_\text{coh}$) for several carbon monolayers. The comparison shows that the cohesive energy ($E_\text{coh}$) of $\beta$-naphthyne (-7.25 eV/atom) is competitive with other 2D carbon allotropes, being close to that of T-graphene (-7.45 eV/atom) and graphenylene (-7.33 eV/atom), but lower than that of pure graphene (-8.02 eV/atom). The $E_\text{coh}$ reflects the energy required to disassemble a solid into isolated atoms, serving as a key indicator of the structural stability of the material. Therefore, while $\beta$-naphthyne shows slightly weaker cohesion than graphene, its value remains sufficiently high to imply a stable 2D structure.

\begin{table*}[pos=htb!]
\centering
\caption{Lattice parameters ($a$ and $b$ in \r{A}, $\alpha = \beta$, and $\gamma$ in degrees), space group (SPG), band gap energy ($E_\text{gap}$) in eV, and cohesive energy ($E_\text{coh}$) in eV/atom obtained for Holey-graphenylene and other relevant 2D carbon allotropes at DFT/GGA-PBE level. The values reported in this table were calculated in this study.} 
\label{tab:properties_comparison} % Rótulo para referência
\begin{tabular}{llllllll}
\hline
  & $a$ & $b$ & $\alpha = \beta$ & $\gamma$ & SPG & $E_\text{gap}$ & $E_\text{coh}$ \\
\hline
$\beta$-naphthyne (this work)   & 10.74 & 11.43 & 90 & 90  & $Cmm$ (65)    & Metallic      & -7.25  \\
T-graphene \cite{sheng2011t}        & 3.45 & 3.45  & 90 & 90 & $P4/mmm$ (123) & Metallic      & -7.45  \\
Twin-graphene \cite{jiang2017twin}     & 6.14 & 6.14  & 90 & 120 & $P6/mmm$ (191) & 0.73 & -7.08  \\
PHE-graphene \cite{zeng2019new}   & 5.73 & 5.73  & 90 & 120 & $P\overline{6}m2$ (187)   & Metallic      & -7.56  \\
Penta-graphene \cite{zhang2015penta} & 3.64 & 3.64 & 90 & 90 & $P42_1m$ (113) & 2.19 & -7.13 \\
Graphyine \cite{narita1998optimized}  & 9.46  & 9.46  & 90 & 120 & $P6/mmm$ (191)    & Metallic & -7.20 \\
Graphenyldiene \cite{laranjeira2024graphenyldiene}     & 6.07 & 6.07  & 90 & 90  & $P4/mbm$ (127)  & 0.78      & -6.92  \\
Graphenylene \cite{song2013graphenylene, du2017new}     & 6.77 & 6.77  & 90 & 120  & $P6/mmm$ (191)  & 0.04      & -7.33  \\
Biphenylene \cite{fan2021biphenylene} & 3.80 & 4.50 &  90 & 90 & $Pmmm$ (47) & Metallic & -7.47 \\
Graphene \cite{geim2007rise}     & 2.47 & 2.47  & 90 & 120  & $P6/mmm$ (191)  & Metallic      & -8.02  \\
\hline
\end{tabular}
\end{table*}

To evaluate the dynamical stability of $\beta$-naphthyne, the phonon band dispersion was computed along the high-symmetry paths of the Brillouin zone, as depicted in Fig. \ref{fig:system}(b). The phonon spectrum confirms the absence of imaginary frequencies, indicating that the material is dynamically stable. At the $\Gamma$-point, the phonon dispersion exhibits three acoustic modes with a characteristic linear behavior. Several phonon branches remain relatively flat in the 18–25 THz range, suggesting low group velocity and localized vibrational modes. In contrast, the bands display quadratic dispersion in the 25–35 THz range, indicating high phononic mobility. Finally, vibrational modes with flat characteristics above 60 THz correspond to the sp-hybridized bonds in $\beta$-naphthyne, a common feature in graphyne-like systems \cite{jenkins2024thd}.

\begin{figure*}[pos=!htb]
    \centering
    \includegraphics[width=1\linewidth]{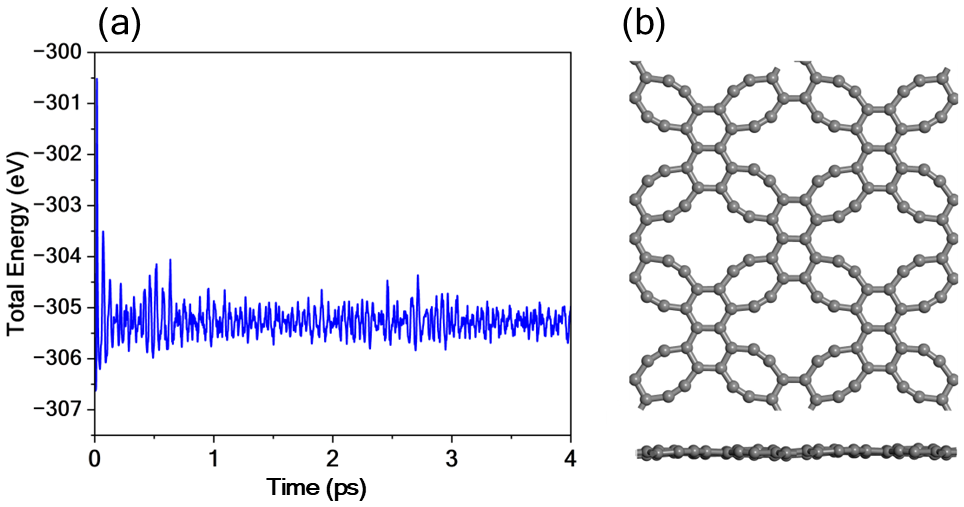}
    \caption{(a) Evolution of total energy as a function of time obtained at 300 K over 4 ps. (b) Final atomic configuration after the AIMD run, showing top and side views of the structure.}
    \label{fig:md}
\end{figure*}

One critical aspect in assessing the viability of novel 2D materials is their thermal stability under ambient conditions. In this way, AIMD simulations were performed at 300 K for a total duration time of 4 ps. Fig. \ref{fig:md}(a) depicts the evolution of the total energy fluctuations during the simulation, along with the final atomic configuration (Fig. \ref{fig:md}(b)).

Fig. \ref{fig:md}(a) shows the total energy as a function of time, with an initial stabilization period within the first picosecond of the simulation. After this phase, energy fluctuations remain minimal, consistently within a range of less than 1 eV, indicating structural robustness. The absence of significant deviations suggests that no phase transitions or reconstruction events occur during the simulation, further supporting the stability of the material at room temperature. The final atomic configuration, presented in Fig. \ref{fig:md}(b), further corroborates this observation. The top and side views reveal that the structural integrity of $\beta$-naphthyne is preserved, with only minor out-of-plane distortions emerging. 

\subsection{Electronic Analysis}
\begin{figure}[pos=!htb]
    \centering
    \includegraphics[width=1\linewidth]{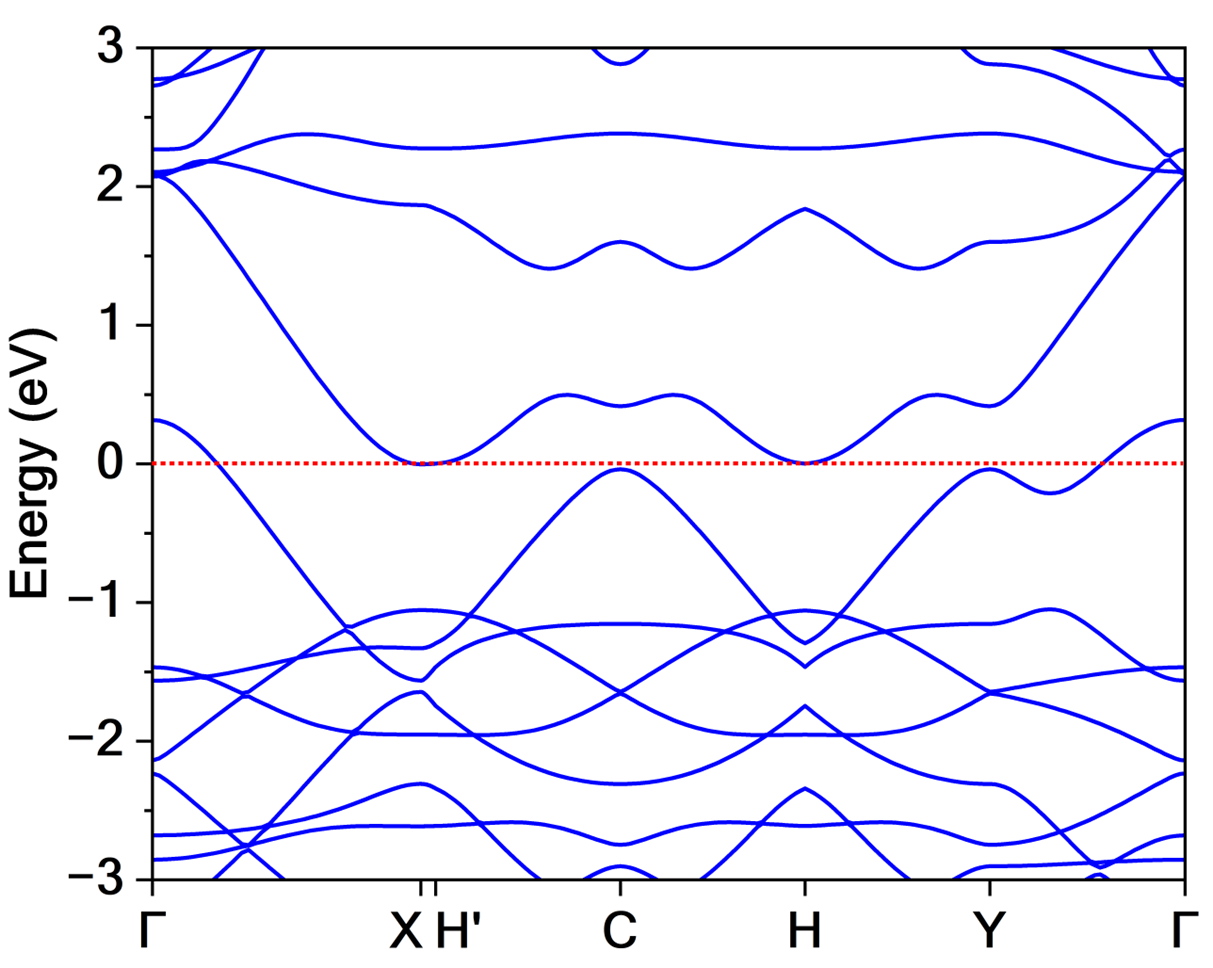}
    \caption{The electronic band structure of $\beta$-naphthyne is depicted along the high-symmetry trajectories within the Brillouin zone. For clarity, the Fermi level is indicated by the red dashed line.}
    \label{fig:band}
\end{figure}

We performed a detailed analysis of the band structure to investigate the electronic properties of $\beta$-naphthyne. The results, depicted in Fig. \ref{fig:band}, confirm that $\beta$-naphthyne exhibits metallic behavior with a band partially occupied crossing the Fermi level ($E_F$) closer to the $\Gamma$ point. The states that compose this band can act as acceptor states, receiving the electronic density and opening the material band gap. This can be useful for an effective interaction with metal adatoms. In the valence band (VB), one can note that the bands display high dispersion, with several crossings, revealing a strong coupling between the electronic states. On the other hand, when the conduction band (CB) is analyzed, the lowest energy band touches $E_F$ two times at the points $H'$ and $H$. This band presents a high dispersion, around 2 eV.

\begin{figure}[pos=!htb]
    \centering
    \includegraphics[width=1\linewidth]{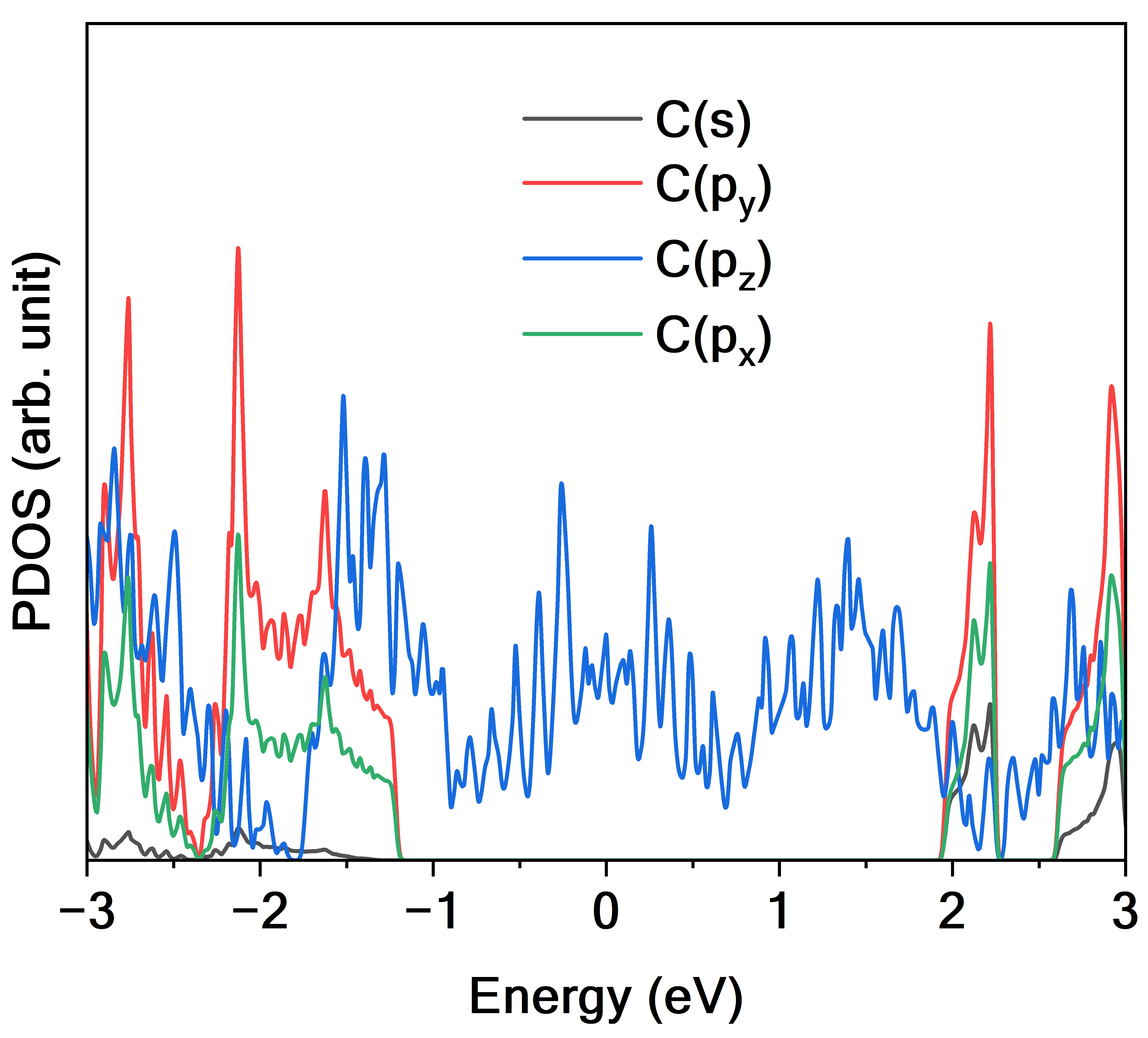}
    \caption{Projected density of states (PDOS) of $\beta$-naphthyne, showing the orbital-resolved contributions to the electronic structure.}
    \label{fig:dos}
\end{figure}

The projected density of states (PDOS) analysis provides a more detailed understanding of the electronic structure, shown in Fig. \ref{fig:dos}, which decomposes the total density of states into atomic orbital contributions. The valence and conduction bands predominantly comprise $p$ orbitals, with the $p_y$ states contributing most significantly, closely followed by the $p_x$ states. The similar energy distribution profiles of the $p_y$ and $p_x$ orbitals suggest a strong orbital coupling in these energy regions. However, near the Fermi level, it becomes evident that only the $p_z$ orbitals contribute, indicating a pronounced $\pi$-character in this energy range.

\begin{figure}[pos=!htb]
    \centering
    \includegraphics[width=1\linewidth]{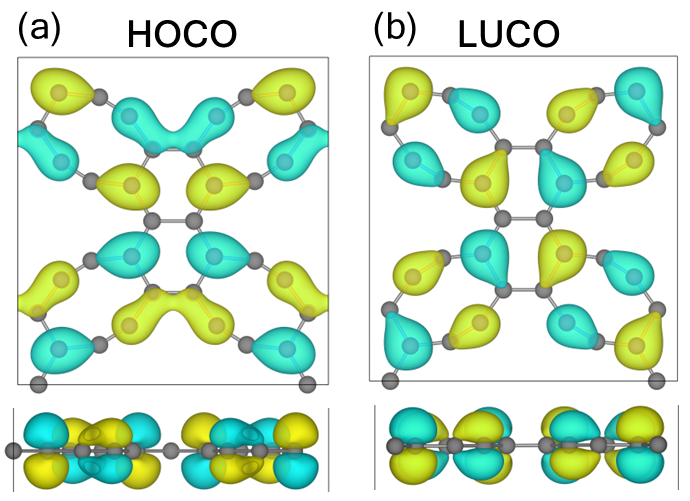}
    \caption{(a) Spatial distribution of the highest occupied crystalline orbital (HOCO) and (b) the lowest unoccupied crystalline orbital (LUCO) of $\beta$-naphthyne.}
    \label{fig:hoco_luco}
\end{figure}

To further investigate the electronic properties of $\beta$-naphthyne, we analyzed the spatial distribution of its frontier molecular orbitals, specifically the highest occupied crystalline orbital (HOCO) and the lowest unoccupied crystalline orbital (LUCO), as shown in Fig. \ref{fig:hoco_luco}. Fig. \ref{fig:hoco_luco}(a) illustrates the HOCO, which is predominantly composed of $\pi$-type orbitals localized on the naphthyl units. The electron density is highly delocalized across the hexagonal rings, indicating a strong conjugation network that enhances hole mobility. The side view further confirms the planarity of the electronic distribution, reinforcing the 2D nature of the conduction pathways.

Fig. \ref{fig:hoco_luco}(b) shows the LUCO, which also exhibits a $\pi$-character, but with electron density more prominently localized around the octagonal rings. This spatial separation between the HOCO and LUCO suggests a moderate charge transfer character, which could influence optical excitations and recombination rates. The LUCO distribution indicates that electron transport might be primarily facilitated through the octagonal framework, complementing the hole conduction pathways of the HOCO. 

The distinct spatial location of HOCO and LUCO further supports the metallic nature of $\beta$-naphthyne, as both orbitals maintain a significant density near the Fermi level. The overlap of these orbitals could also contribute to strong optical absorption, suggesting potential applications in photodetectors and optoelectronic devices, as will be discussed later.

\begin{figure}[pos=!htb]
    \centering
    \includegraphics[width=0.8\linewidth]{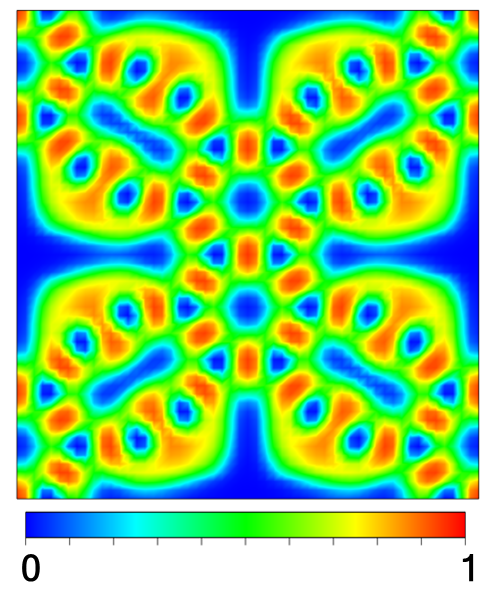}
    \caption{Electron Localization Function (ELF) of $\beta$-naphthyne.}
    \label{fig:ELF}
\end{figure}

To gain a deeper understanding of the charge distribution and bonding nature of $\beta$-naphthyne, we analyzed its electron localization function (ELF), as shown in Fig. \ref{fig:ELF}. The ELF provides valuable information on the degree of electron localization, where a value of 1 indicates strongly localized electrons (e.g., covalent bonds or lone pairs), a value of 0.5 corresponds to a homogeneous electron gas-like distribution, and a value of 0 represents regions with minimal or no electron density.

In $\beta$-naphthyne, the naphthyl fragments exhibit high ELF values, consistent with the resonant nature of benzene rings, where delocalized $\pi$-electrons contribute to moderately high electron localization due to sp$^{2}$ hybridization. In contrast, the octagonal rings exhibit lower ELF values, indicating greater electron delocalization than the naphthyl units. This behavior can be attributed to sp-hybridized bonds, which result in a more diffuse electron density along the in-plane bonds. The lower ELF in these regions suggests that the electronic structure of the octagonal motifs contributes to anisotropic charge transport, complementing the conduction pathways of the naphthyl fragments.

The ELF map further supports the conclusion that $\beta$-naphthyne exhibits a highly delocalized electronic structure, with distinct regions contributing to different transport properties. The hybridized bonding interactions within the naphthyl and octagonal rings create a unique electronic environment, making this material a promising candidate for nanoelectronic and optoelectronic applications, where controlled charge delocalization and transport are essential.

\subsection{Optical Properties}

\begin{figure}[pos=!htb]
    \centering
    \includegraphics[width=\linewidth]{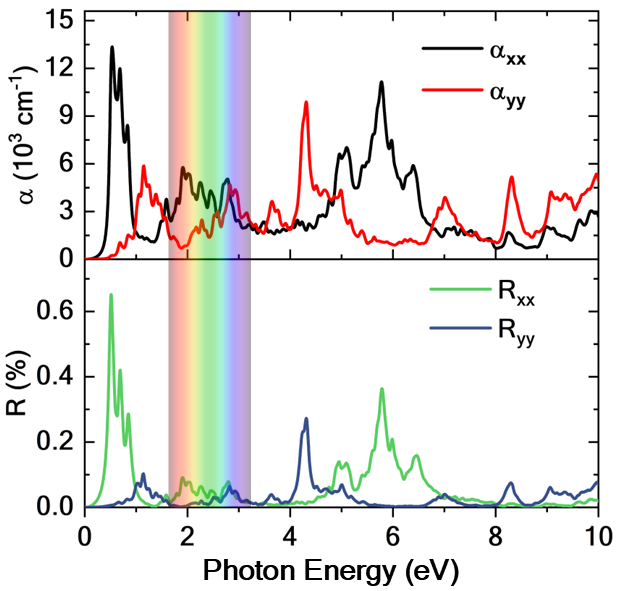}
    \caption{Optical properties of $\beta$-naphthyne. (Top panel) Absorption coefficient ($\alpha$) as a function of photon energy for the $x$ (black) and $y$ (red) polarization components. (Bottom panel) Reflectance ($R$) for the $x$ (green) and $y$ (blue) components.}
    \label{fig:optical}
\end{figure}

To address the optical properties of $\beta$-naphthyne, we calculated its absorption coefficient ($\alpha$) and reflectance ($R$), as illustrated in Fig. \ref{fig:optical}. The visible light range is highlighted in the figure to facilitate the identification of its transparency and absorption characteristics.

$\beta$-naphthyne exhibits strong absorption in the IR region, where the absorption coefficient reaches approximately $14 \times 10^3$ cm$^{-1}$ for the $x$ component. This result suggests that the material could be well-suited for IR photodetectors and thermal shielding applications. Additionally, the $y$ component shows noticeable absorption in the near-IR range, further reinforcing its potential use in IR-active optoelectronic devices. 

The material displays relatively low absorption in the visible spectrum, with the maximum absorption coefficient reaching $6 \times 10^3$ cm$^{-1}$ between the red and orange spectral regions for the $x$ component. This trend indicates that $\beta$-naphthyne is mainly transparent to visible light, making it a promising candidate for transparent conducting films or low-energy-loss optical coatings. The reflectance in this range remains minimal, further confirming that the material has low optical reflection and high light transmission properties. $\beta$-naphthyne exhibits strong absorption peaks in the UV region, with significant transitions occurring around 4.5 eV ($y$ component) and 6 eV ($x$ component). These absorption peaks are accompanied by increased reflectance, indicating that the material effectively interacts with high-energy photons in this range.

\subsection{Vibrational Properties}

\begin{figure}[pos=!htb]
    \centering
    \includegraphics[width=1\linewidth]{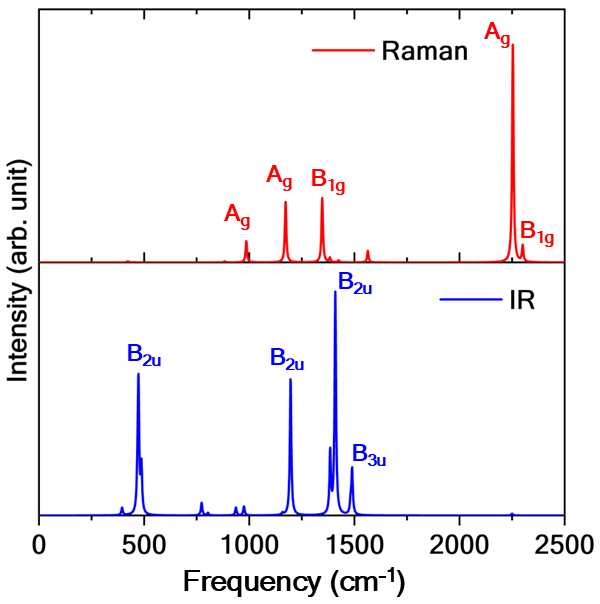}
    \caption{(Top/Bottom panel) Raman/Infrared spectrum of $\beta$-naphthyne.}
    \label{fig:Raman+IR}
\end{figure}

Vibrational spectroscopy provides essential insights into the bonding nature and structural integrity of materials. To assess the phonon behavior of $\beta$-naphthyne, we computed both the Raman and IR spectra, as shown in Fig. \ref{fig:Raman+IR}. These spectra allow us to distinguish between Raman-active and IR-active vibrational modes, shedding light on the symmetry and electronic environment of atomic vibrations in the material.

To obtain Raman and IR spectra (Fig.  \ref{fig:Raman+IR}), the CPHF/KS approach implemented in CRYSTAL17 was utilized. The vibration modes of holey-naphthylene can be represented by $\Gamma_{vib}$ = 5$B_{1u}$ + 9$B_{2u}$ + 9$B_{3u}$ + 4$B_{2g}$ + 4$A_{u}$ + 5$B_{3g}$ + 9$B_{1g}$ + 9$A_{g}$. Between these modes, three are acoustic ($B_{1u}$ + $B_{2u}$ + $B_{3u}$), four are silent (4$A_{u}$), 23 are IR-active (5$B_{1u}$ + 9$B_{2u}$ + 9$B_{3u}$), and 27 are Raman-active (4$B_{2g}$ + 5$B_{3g}$ + 9$B_{1g}$ + 9$A_{g}$).

Analyzing the Raman spectrum (Fig. \ref{fig:Raman+IR}, top panel) reveals several distinct peaks, with the most intense modes $Ag$ and $B_{1g}$ occurring at low and intermediate frequencies. Four prominent bands are visible at 985 cm$^{-1}$ ($A_{g}$), 1172 cm$^{-1}$ ($A_{g}$), 1383 cm$^{-1}$ ($B_{1g}$), and 2253 cm$^{-1}$ ($A_{g}$). Strong $A_g$ symmetric modes suggest collective in-plane stretching and breathing vibrations of the carbon framework. The sharp intensity of these peaks highlights the high rigidity of the system, which is commonly observed in sp$^{2}$-hybridized carbon networks. The prominent peak at the high-frequency region corresponds to a strong C--C stretching mode, a signature of a well-preserved conjugated structure in the material.

However, by verifying the IR spectrum (Fig. \ref{fig:Raman+IR}, bottom panel), several intense peaks associated with asymmetric stretching and bending modes are shown. Four intense bands are noticed at 472 cm$^{-1}$ ($B_{2u}$), 1196 cm$^{-1}$ ($B_{2u}$), 1409 cm$^{-1}$ ($B_{2u}$), and 1489 cm$^{-1}$ ($B_{3u}$) are observed. The $B_{2u}$ and $B_{3u}$ indicate dipole-active vibrations arising from charge asymmetry within the structure. The strongest IR peak, located around 1500 cm$^{-1}$, is related to C--C stretching motions similar to those observed in other 2D carbon-based materials \cite{esposito2024anharmonic, kuznetsov2022dependence}. The presence of multiple low-frequency peaks suggests the existence of out-of-plane bending and deformation modes, which contribute to the dynamic flexibility of the monolayer.

\subsection{Mechanical Properties}

The mechanical properties of $\beta$-naphthyne were assessed by calculating its elastic constants, which are crucial to determining the material's structural stability and mechanical response. The obtained values for the independent elastic constants are $C_{11} = 169.10$ N/m, $C_{22} = 173.64$ N/m, $C_{12} = 42.68$ N/m, $C_{16} = -22.32$ N/m, $C_{26} = -17.49$ N/m, and $C_{66} = 20.98$ N/m. According to the Born-Huang stability criteria \cite{born1940stability}, the mechanical stability of 2D oblique systems is ensured if $C_{11} > 0$, $C_{11}C_{22} > C_{12}^2$, and $\det(C_{ij}) > 0$. Substituting the calculated values, these conditions are all satisfied. Therefore, we confirm that the $\beta$-naphthyne structure is mechanically stable under small deformations.

\begin{figure*}[pos=!htb]
    \centering
    \includegraphics[width=1\linewidth]{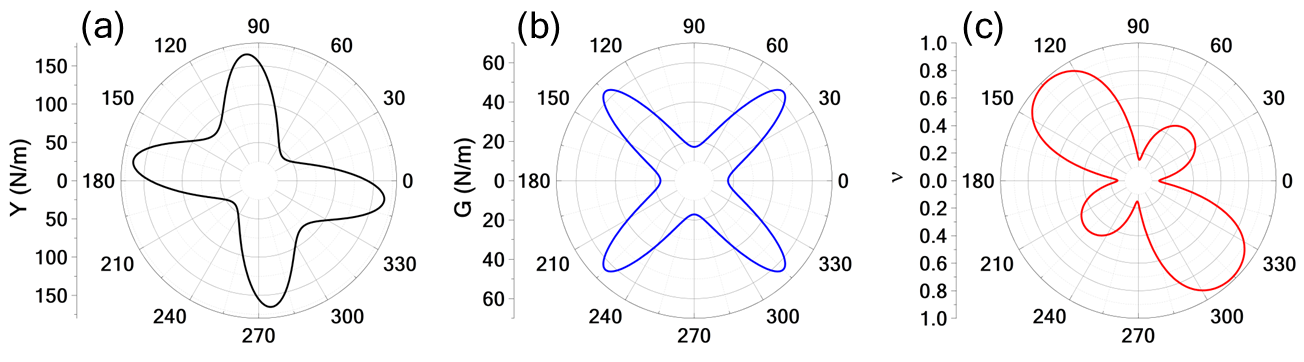}
    \caption{Polar plots of (a) Young’s modulus ($Y$), (b) Shear modulus ($G$), and (c) Poisson's ratio ($\nu$).} for $\beta$-naphthyne, demonstrating their mechanical anisotropy.
    \label{fig:polar}
\end{figure*}

One can note that Young's modulus ($Y$), which quantifies the material's resistance to uniaxial deformation, exhibits strong anisotropy. As shown in Fig. \ref{fig:polar}(a), $Y$ reaches a maximum of 165.88 N/m along one crystallographic direction and a minimum of 43.71 N/m in the perpendicular direction, resulting in an anisotropy ratio of 3.79. This trend indicates that the material's stiffness strongly depends on the applied strain direction. The maximum Young's modulus of $\beta$-naphthyne surpasses that of graphenyldiene (122.47 N/m) and graphyne (123.59 N/m) (refer to Table \ref{tab:elastic_constants}), reinforcing its classification as a graphyne-like monolayer due to the presence of sp-bonded carbon atoms.

\begin{table*}[pos=!htb]
\centering
\caption{Maximum and minimum values of Young's modulus (\(Y_{\text{max}}, Y_{\text{min}}\)) (N/m), Poisson ratio (\(\nu_{\text{max}}, \nu_{\text{min}}\)), and Shear modulus (\(G_{\text{max}}, G_{\text{min}}\)) (N/m) for $\alpha$, $\beta$, and $\gamma$-anthraphenylenes and other carbon monolayers. The values reported in this table were calculated in this study.}
\begin{tabular}{lccc}
\hline
 & \(Y_{\text{max}}/Y_{\text{min}}\) & \(\nu_{\text{max}}/\nu_{\text{min}}\) & \(G_{\text{max}}/G_{\text{min}}\) \\
\hline
$\beta$-naphthyne  & 165.88/43.71 & 0.98/0.15 & 63.94/17.17 \\ 
PHE-graphene & 262.29/262.29 & 0.26/0.26 & 103.91/103.91 \\
Graphenylene & 209.02/209.02 & 0.27/0.27 & 82.11/82.11 \\
Graphene & 345.42/345.42 & 0.17/0.17 & 147.60/147.60 \\
Graphenyldiene & 122.47/122.47 & 0.35/0.35 & 45.29/45.29 \\
Penta-graphene & 271.81/266.67 & -0.08/-0.10 & 151.21/144.98 \\
T-graphene & 293.90/148.02 & 0.16/0.58 & 126.57/148.02 \\ 
Graphyne & 123.59/123.59 & 0.45/0.45 & 42.63/42.63 \\
\hline
\end{tabular}
\label{tab:elastic_constants}
\end{table*}

The shear modulus ($G$), which describes the resistance to shear deformations, also demonstrates notable directional variation. Fig. \ref{fig:polar}(b) depicts that $G$ varies from 63.93 N/m to 17.17 N/m, corresponding to an anisotropy ratio of 3.73. Interestingly, the maximum shear modulus exceeds that of graphenyldiene (45.29 N/m) and graphyne (42.63 N/m), suggesting that $\beta$-naphthyne resists shear forces more effectively than other related 2D carbon materials. The Poisson’s ratio ($\nu$), shown in Fig. \ref{fig:polar}(c), exhibits the highest degree of anisotropy, ranging from 0.98 to 0.15, with an anisotropy ratio of 6.48. This extreme variation indicates that $\beta$-naphthyne’s ability to contract in one direction when stretched in another highly depends on the strain orientation. Interestingly, the material can mimic Poisson’s graphene ratio, which is considered one of the most mechanically robust materials. The pronounced mechanical anisotropy of $\beta$-naphthyne is among the highest reported for 2D carbon allotropes, even exceeding that of T-graphene, which was previously considered one of the most anisotropic 2D materials.

\subsection{One-dimensional (1D) structures}

\begin{figure*}
    \centering
    \includegraphics[width=1\linewidth]{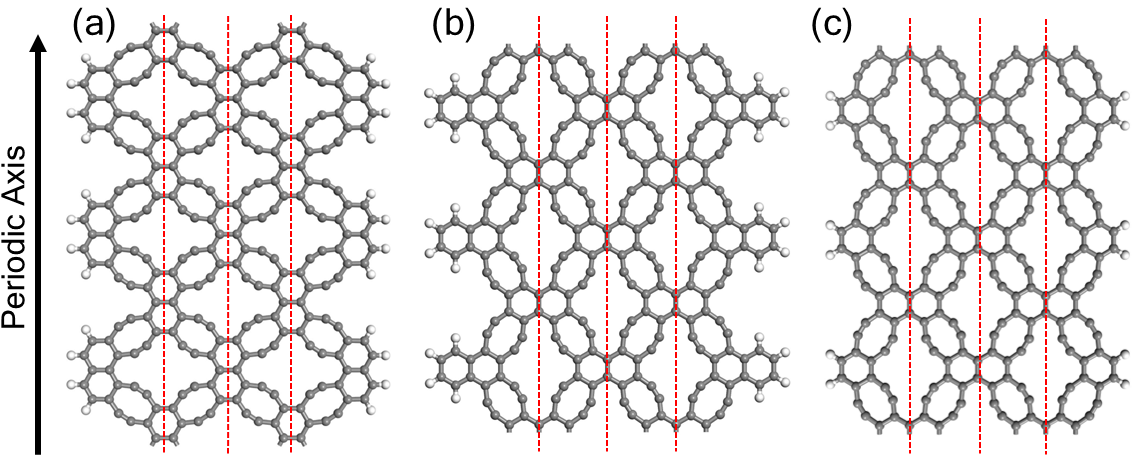}
    \caption{Schematic representation of the 1D structures of $\beta$-naphthyne nanoribbons. (a) Zigzag nanoribbon (ZNNR), (b) armchair nanoribbon with double hexagon edges (ANNR-$hh$), and (c) armchair nanoribbon with single hexagon edges (ANNR-$h$).}
    \label{fig:ribbons}
\end{figure*}

We now turn our attention to nanoribbons derived from $\beta$-naphthyne. Due to their rectangular lattice, the nanoribbons were generated by exploring the cutting directions of the armchair ($y$ or $b$) and zigzag ($x$ or $a$). The armchair and zigzag $\beta$-naphthyne nanoribbons are referred to as ANNR and ZNNR, respectively. For ZNNR, we considered a single type of edge in the nonperiodic direction, consisting of two hexagons, as illustrated in Fig. \ref{fig:ribbons}a. In the case of ANNRs, two edge terminations were examined: one with two hexagons ($hh$) (Fig. \ref{fig:ribbons}b) and another with a single hexagon ($h$) (Fig. ~\ref{fig:ribbons}c), labeled ANNR-$hh$ and ANNR-$h$, respectively. The chosen nanoribbon width spans three naphthyl groups. This methodology is in agreement with the approach adopted by Paz et al. \cite{ALVARESPAZ2019792} in their investigation of the 1D structures of $\beta$-naphthylene. Note that $\beta$-naphthyne nanoribbons did not show reconstructions or bond breaks in optimization. In addition, the planarity of 2D $\beta$-naphthyne was maintained. 

\begin{figure*}
    \centering
    \includegraphics[width=1\linewidth]{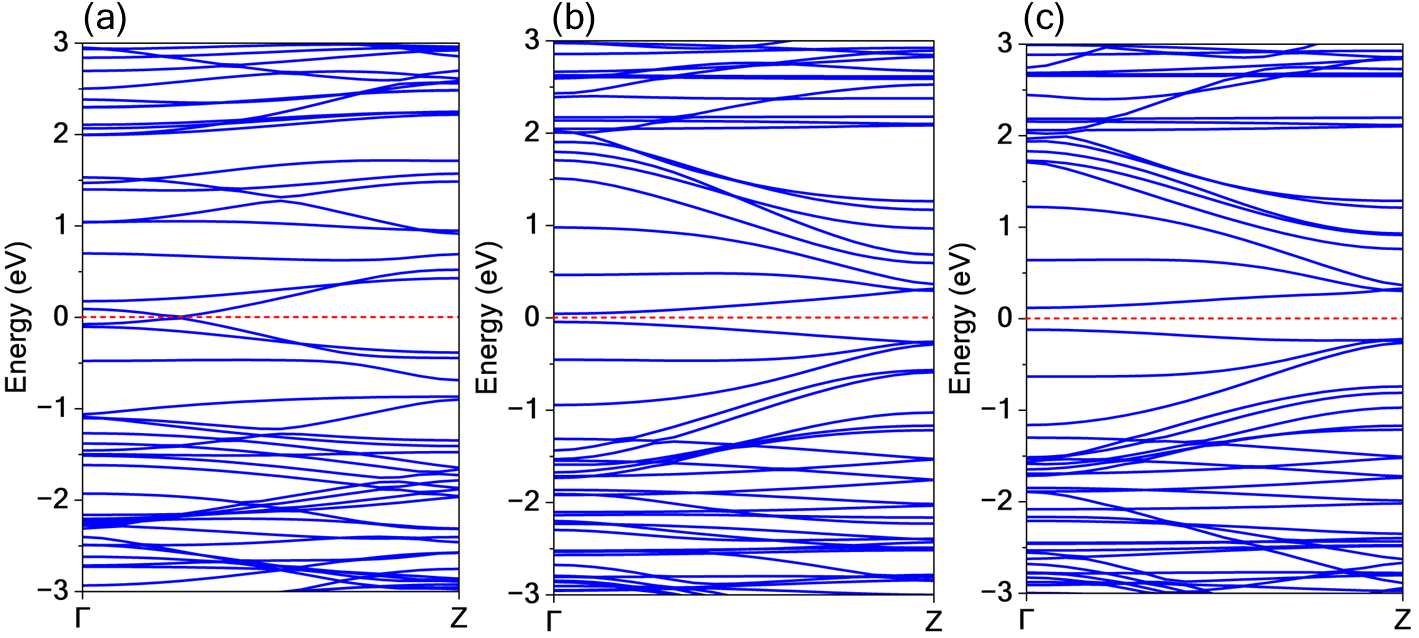}
    \caption{The electronic band structure of $\beta$-naphthyne 1D structures for (a) ZNNR, (b) ANNR-$hh$, and ANNR-$h$. The red dashed line indicates the Fermi level.}
    \label{fig:band_ribbon}
\end{figure*}

To explore the electronic properties of the $\beta$-naphthyne 1D structures, we calculated the band structures for ZNNR, ANNR-$h$, and ANNR-$hh$, as depicted in Fig.  \ref{fig:band_ribbon}. The ZNNR retains a metallic character, mirroring its 2D counterpart. However, a remarkable characteristic emerges with a Dirac cone positioned closer to the $\Gamma$ point. In contrast, the ANNR-$hh$ and ANNR-$h$ structures reveal a transition from metallic to semiconducting behavior, with direct band gaps at the $\Gamma$ point of 0.08 eV and 0.24 eV, respectively. This band gap opening can be attributed to the edge-induced symmetry-breaking and quantum confinement effects inherent to the reduced dimensionality.

\begin{figure*}
    \centering
    \includegraphics[width=1\linewidth]{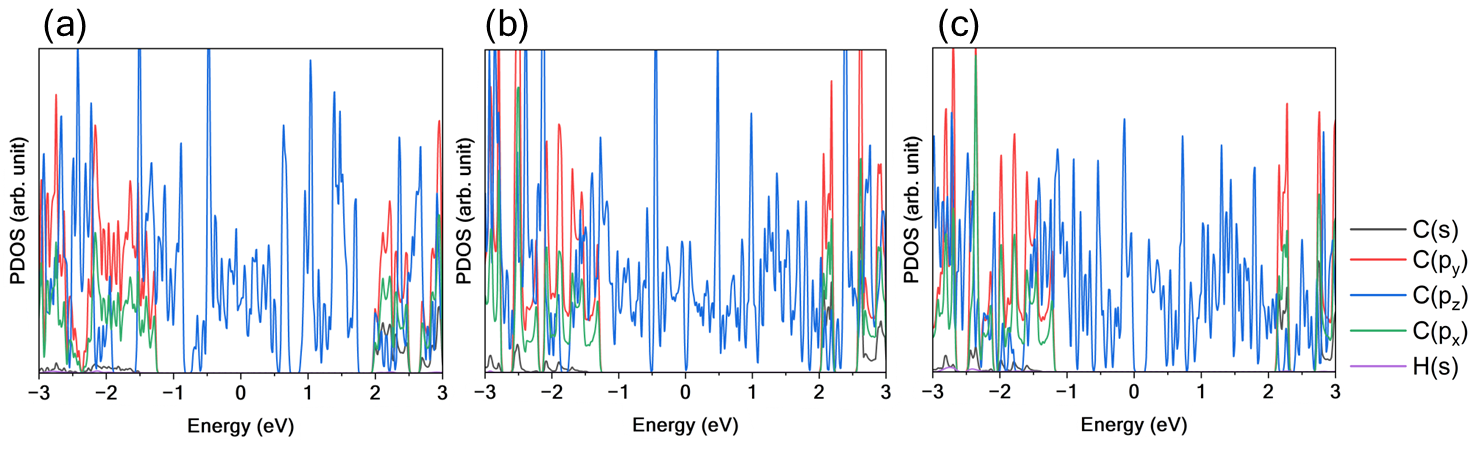}
    \caption{Projected density of states (PDOS) of $\beta$-naphthyne 1D structures for (a) ZNNR, (b) ANNR-$hh$, and ANNR-$h$.}
    \label{fig:dos_ribbon}
\end{figure*}

To verify the composition of the electronic states of the $\beta$-naphthyne nanoribbons, the projected density of states (PDOS) was calculated, as shown in Fig.  \ref{fig:dos_ribbon}. The general features closely resemble those of the 2D counterpart, such as the prominent contribution of $p_y$ and $p_x$ states at the lower and higher energies of the valence band (VB) and conduction band (CB), respectively. Around the Fermi level ($E_F$), the $p_z$ states dominate entirely, with no significant contributions from other orbitals. Furthermore, hydrogen states do not show relevant influence across the evaluated energy range.

For ZNNR (Fig.  \ref{fig:dos_ribbon}(a)), despite the presence of a Dirac cone, the PDOS reveals a non-zero density of states at the Fermi level. This finite DOS arises from the low dispersion of bands near the Dirac point, causing an accumulation of electronic states around $E_F$, preventing the DOS from vanishing completely. In the cases of ANNR-$hh$ (Fig.  \ref{fig:dos_ribbon}(b)) and ANNR-$h$ (Fig.  \ref{fig:dos_ribbon}(c)), a noticeable accumulation of states at the band edges is observed, accompanied by an abrupt drop in the DOS at these regions, characteristic of highly correlated electronic states. 

\section{Conclusion}

In this work, we explore the structural, electronic, mechanical and optical properties of a novel porous 2D carbon allotrope, $\beta$-naphthyne, using first-principles calculations. Our results confirmed its stability, with phonon dispersion and AIMD simulations indicating a robust framework.

The electronic band structure revealed that $\beta$-naphthyne exhibits metallic behavior, with strong $\pi$-characther for the states in the band edges. Mechanical analysis highlighted its strong anisotropy, showing a significant directional dependence on Young's modulus, shear modulus, and Poisson’s ratio. This mechanical anisotropy surpasses that of previously studied 2D carbon materials, positioning $\beta$-naphthyne as a promising candidate for strain engineering applications.

Optical investigations demonstrated that $\beta$-naphthyne exhibits strong IR absorption, moderate interaction within the visible spectrum, and pronounced UV absorption peaks, suggesting its potential for optoelectronic and photonic technologies. Furthermore, the Raman and IR spectra provided distinctive vibrational fingerprints, providing a reliable reference for future experimental validation.

We further examine the electronic properties of 1D $\beta$-naphthyne nanoribbons. Although ZNNR retained the metallic character of its 2D counterpart, ANNR-$h$ and ANNR-$hh$ showed a transition from metallic to semi-conducting behavior, with direct band gaps of 0.08 and 0.24 eV, respectively. These findings underscore the tunable electronic properties of $\beta$-naphthyne, driven by dimensionality reduction and edge geometry, broadening its potential for nanoelectronic applications.

\section*{Data access statement}
Data supporting the results can be accessed by contacting the corresponding author.

\section*{Conflicts of interest}
The authors declare no conflict of interest.

\section*{Acknowledgements}
This work was supported by the Brazilian funding agencies Fundação de Amparo à Pesquisa do Estado de São Paulo - FAPESP (grant no. 2022/03959-6, 2022/14576-0, 2020/01144-0, 2024/05087-1, and 2022/16509-9), and National Council for Scientific, Technological Development - CNPq (grant no. 307213/2021–8). L.A.R.J. acknowledges the financial support from FAP-DF grants 00193.00001808/2022-71 and $00193-00001857/2023-95$, FAPDF-PRONEM grant 00193.00001247/2021-20, PDPG-FAPDF-CAPES Centro-Oeste 00193-00000867/2024-94, and CNPq grants $350176/2022-1$ and $167745/2023-9$. 

\printcredits

\bibliography{biblio/cas-refs}

\end{document}